\documentclass[12pt,a4paper]{article}
\usepackage{a4wide,cite}
\usepackage{amsmath,amssymb}
\usepackage{latexsym,graphicx,epsfig}

\def\lsim{\mathrel{\rlap{\raise 2.5pt \hbox{$<$}}\lower 2.5pt\hbox{$\sim$}}}
\def\gsim{\mathrel{\rlap{\raise 2.5pt \hbox{$>$}}\lower 2.5pt\hbox{$\sim$}}}

\allowdisplaybreaks 

\begin{document}
\renewcommand{\thefootnote}{\fnsymbol{footnote}}
\newpage\normalsize
    \pagestyle{plain}
    \setlength{\baselineskip}{4ex}\par
    \setcounter{footnote}{0}
    \renewcommand{\thefootnote}{\arabic{footnote}}
\newcommand{\preprint}[1]{%
\begin{flushright}
\setlength{\baselineskip}{3ex} #1
\end{flushright}}
\renewcommand{\title}[1]{%
\begin{center}
    \LARGE #1
\end{center}\par}
\renewcommand{\author}[1]{%
\vspace{2ex}
{\Large
\begin{center}
    \setlength{\baselineskip}{3ex} #1 \par
\end{center}}}
\renewcommand{\thanks}[1]{\footnote{#1}}
\begin{flushright}
    \today
\end{flushright}
\vskip 0.5cm

\begin{center}
{\bf \Large Top and Bottom Squarks Decays under Cosmological Bounds}
\end{center}
\vspace{1cm}
\begin{center}
  L. Selbuz
\footnote{e-mail address: selbuz@eng.ankara.edu.tr }
    and
  Z. Z. Aydin
  \footnote{e-mail address: zzaydin@eng.ankara.edu.tr
 }
  \end{center}
\vspace{1cm}
\begin{center}
 Department of Engineering Physics, Faculty of Engineering,
Ankara University, \\
06100  Tandogan-Ankara, Turkey
\end{center}
\vspace{1cm}
\begin{abstract}
We investigate the fermionic decays of top squarks $\tilde t_{1,2}$
and bottom squarks $\tilde b_{1,2}$ in the Minimal Supersymmetric
Standard Model with complex parameters $M_1$, $\mu$, $A_t$ and $A_b$
. In the analysis we particularly take into account the cosmological
bounds imposed by WMAP data. We plot the CP phase dependences of
stop and sbottom decay widths.
\end{abstract}


\section{Introduction}
\setcounter{equation}{0}
SUSY provides an appealing realization of the Higgs mechanism for
mass generation. When SUSY is broken softly, superpartners acquire
masses not exceeding 1 TeV. Since these new particles are in the
exploration range of the LHC, we have to analyse non-standard Higgs
sector extensively in all possible ways \cite{Haber}.

The minimal supersymmetric extension of the Standard Model (MSSM)
requires a non-minimal Higgs sector \cite{higgssector} which
introduces additional sources of CP-violation beyond the economical
Kobayashi-Maskawa phase of the SM.  The plethora of CP-phases also
influences the decays and mixings of B mesons (as well as D and K
mesons). The present experiments at BABAR, Tevatron and KEK and the
one to start at the LHC will be able to measure various decay
channels to determine if there are supersymmetric sources of CP
violation. In particular, CP-asymmetry and decay rate of $B
\rightarrow X_s \gamma$ form a good testing ground for low-energy
supersymmetry with CP violation \cite{bmeson}. These additional
sources of CP-violation are welcome to explain the cosmological
baryon asymmetry of the universe. In addition to this, the lightest
superpartner, i.e. the lightest neutralino $\tilde \chi_1^0 $ could
be an excellent candidate for cold dark matter in the universe.

With the precision experiments by Wilkinson Microwave Anisotropy
Probe (WMAP), the relic density of cold dark matter can be
constrained to $0.0945 < \Omega_{CDM} h^2 < 0.1287$ at $2\sigma$
level \cite{Wilkinson}. Recently, in the light of this cosmological
constraint, an extensive analysis of the neutralino relic density in
the presence of CP phases has been given by B\'{e}langer  {\it et
 al.} \cite{Belanger}.

Analyses of the decays of third generation scalar quarks with
complex SUSY parameters have been performed by Bartl et al.
\cite{Bartl}. In the present note we repeat this analysis taking
into account the cosmological bound imposed by WMAP. Namely, we
study the effect of $M_1$ and its phase $ \varphi_{U(1)}$ on the
decay widths of $\tilde t_{1,2}$ and $\tilde b_{1,2}$. In the
numerical calculations, although the SUSY parameters $\mu$, $M_1$,
$M_2$, and $A_f$ are in general complex, we assume that $\mu$,
$M_2$, $A_t$ and $A_b$ are real, but $M_1$ and its phase $
\varphi_{U(1)}$ take values on the WMAP- allowed bands given in Ref.
\cite{Belanger}. These bands also satisfy the EDM bounds
 \cite{edms}. And we evaluate the parameter $M_2$ via the relation $
M_2=3/5|M_1|(\tan\theta_W)^{-2}$. It is very important to insert the
WMAP-allowed band in the plane $ M_1-\varphi$ into the numerical
calculations instead of taking one fixed $M_1$ value for all
$\varphi$-phases, because, for example, on the allowed band for
$\mu=200$ GeV, $M_1$ starts from 140 GeV for $\varphi=0$ and
increasing monotonously it becomes 165 GeV for $\varphi=\pi$. In
Ref.\cite{Belanger} two WMAP-allowed band plots are given, one for
$\mu=200$ GeV and the other for $\mu=350$ GeV. For both plots the
other parameters are fixed to be $\tan \beta=10$, $ m_{H^+}=1$ TeV,
$A_t=1.2 $ TeV, $A_b=1.2 $ TeV,
$\varphi_\mu$=$\varphi_{A_t}$=$\varphi_{A_b}$=0.


\section{Top and Bottom Squarks Masses, Mixing and Decay Widths }
\setcounter{equation}{0}

\subsection{Masses and mixing in squark sector}
The superpartners of the SM fermions with left and right helicity
are the left and right sfermions. In the case of top squark (stop)
and bottom squark (sbottom) the left and right states are in general
mixed. Therefore, the sfermion mass terms of the Lagrangian are
described in the basis ($\tilde q_{L}$,$\tilde q_{R}$) as
\begin{equation}\label{mkl1}
 {\cal L}_M^{\tilde q }= -({\tilde q}_L^{\dag}{\tilde q}_R^{\dag})\left(
 \begin{array}{cc}
 M_{L L}^{2}& M_{L R}^{2}\\[1.ex]
 M_{R L}^{2} & M_{R R}^{2}
 \end{array}
 \right)
 \left(
\begin{array}{c}
\tilde q_L\\ [1.ex] \tilde q_R
\end{array}
\right)
\end{equation}
with
\begin{eqnarray}\label{mkl2}
M_{L L}^{2}&=&M_{\tilde Q}^{2}+(I_{3L}^{q}-e_q\sin^2\theta_W)\cos(2
\beta)m_{z}^{2}+m_{q}^{2}\\
M_{R R}^{2}&=&M_{\tilde Q'}^{2}+e_q\sin^2\theta_W\cos(2
\beta)m_{z}^{2}+m_{q}^{2}\\\label{mkl3} M_{R L}^{2}&=&(M_{L
R}^{2})^{*}=m_q(A_q-\mu^{*}(\tan\beta)^{-2I_{3L}^{q}})\label{mkl4}
\end{eqnarray}
where $m_q$, $e_q$, $I_{3L}^{q}$ and $\theta_W$ are the mass,
electric charge, weak isospin of the quark q=b,t and the weak mixing
angle, respectively. $\tan\beta=v_2/v_1$ with $v_i$ being the vacuum
expectation values of the Higgs fields $H_i^{0}$, $ i=1,2$. The
soft-breaking parameters $M_{\tilde Q}$, $M_{\tilde Q'}=M_{\tilde U}
(M_{\tilde D})$ for q=t(b), $A_b$ and $A_t$ involved in Eqs.
(2.2-2.4) can be evaluated for our numerical calculations using the
following relations
\begin{eqnarray}\label{mkl5}
M_{\tilde Q}^{2}&=&\frac{1}{2}{\left(m_{\tilde t_1}^{2}+m_{\tilde
t_2}^{2} \pm\sqrt{(m_{\tilde t_2}^{2}-m_{\tilde t_1}^{2})^2-4m_t^{2}
|A_t-\mu^{*}\cot\beta|^2}\right)}\nonumber \\
&&-(\frac{1}{2}-\frac{2}{3}\sin^2\theta_W)\cos(2\beta)m_{z}^{2}-m_{t}^{2}\\
\label{mkl6} M_{\tilde U}^{2}&=&\frac{1}{2}{\left(m_{\tilde
t_1}^{2}+m_{\tilde t_2}^{2} \mp\sqrt{(m_{\tilde t_2}^{2}-m_{\tilde
t_1}^{2})^2-4m_t^{2}
|A_t-\mu^{*}\cot\beta|^2}\right)}\nonumber \\
&&-\frac{2}{3}\sin^2\theta_W\cos(2\beta)m_{z}^{2}-m_{t}^{2}
\end{eqnarray}
and similar ones for $M_{\tilde Q}$ and $M_{\tilde D}$ by
interchanging $t\leftrightarrow b$ in Eqs. (2.5-2.6 ).

The squark mass eigenstates $\tilde q_1$ and $\tilde q_2$
can be obtained from the weak states $\tilde q_L$ and $\tilde q_R$ via the $\tilde q$-mixing matrix
\begin{equation}\label{mkl10}
 {\cal R}^{\tilde q }=\left(
 \begin{array}{cc}
 e^{i\varphi_{\tilde q}}\cos\theta_{\tilde q}& \sin\theta_{\tilde q}\\[1.ex]
 -\sin\theta_{\tilde q} & e^{-i\varphi_{\tilde q}}\cos\theta_{\tilde q}
 \end{array}
 \right)
\end{equation}
where
 \begin{equation}\label{mkl9}
 \varphi_{\tilde q}=\arg[M_{R
 L}^{2}]=\arg[A_q-\mu^{*}(\tan\beta)^{-2I_{3L}^{q}}]
 \end{equation}
 and
\begin{equation}\label{mkl11}
 \cos\theta_{\tilde q}=\frac{-|M_{L R}^{2}|}
 {\sqrt{|M_{L R}^{2}|^2+
 (m_{\tilde q_1}^{2}-M_{L L}^{2})^2}}, \qquad
\sin\theta_{\tilde q}=\frac{M_{L L}^{2}-m_{\tilde q_1}^{2}}
 {\sqrt{|M_{L R}^{2}|^2+
 (m_{\tilde q_1}^{2}-M_{L L}^{2})^2}}
\end{equation}
One can easily get the following squark mass eigenvalues by diagonalizing the mass matrix in Eq. (2.1):
\begin{equation}\label{mkl12}
 m_{\tilde q_{1,2}}^{2}=\frac{1}{2}
{\left(M_{L L}^{2}+M_{R R}^{2} \mp\sqrt{(M_{L L}^{2}-M_{R
R}^{2})^2+4|M_{L R }^{2}|^2} \right)} ,\qquad m_{\tilde q_1}<
m_{\tilde q_2}
\end{equation}

We might add a comment about the possibility of a flavor mixing, for
example, between the second and third squark families. In this case,
the sfermion mass matrix in Eq. (2.1) becomes a 4x4 matrix in the
basis ($\tilde c_L$, $\tilde c_R$, $\tilde t_L$, $\tilde t_R$). Then
one obtains squark mass eigenstates ($\tilde c_1$, $\tilde c_2$,
 $\tilde t_1$, $\tilde t_2$) from these weak states, and analyzes
their decays by utilizing procedures similar to the ones indicated
in the text. The problem with flavor violation effects is that their
inclusion necessarily correlates B, D and K physics with direct
sparticle searches at colliders. Moreover, it has been shown that,
with sizeable supersymmetric flavor violation, even the Higgs
phenomenology at the LHC correlates with that of the rare processes
\cite{flavor}. In this work we have neglected such effects; however,
we emphasize that inclusion of such effects can give important
information on mechanism that breaks supersymmetry via decay
products of squarks.

\subsection{Fermionic decay widths of  $\tilde t_i $ and  $\tilde b_i $ }
The quark-squark-chargino and quark-squark-neutralino Lagrangians
have been first given in Ref. \cite{Haber}. Here we use them in
notations of Ref. \cite{Bartl}:
\begin{eqnarray}\label{mkl13}
{\cal L}_{q \tilde q \tilde \chi^{+}}=g\bar{t}(\ell_{i j}^{\tilde
b}P_R + k_{i j}^{\tilde b}P_L){\tilde \chi}_j^{+}{\tilde b}_i+g\bar{b}(\ell_{i j}^{\tilde
t}P_R + k_{i j}^{\tilde t}P_L){\tilde \chi}_j^{+c}{\tilde t}_i+h.c.
\end{eqnarray}
and
\begin{eqnarray}\label{mkl18}
{\cal L}_{q \tilde q \tilde \chi^{0}} =g\bar{q}(a_{i k}^{\tilde
q}P_R + b_{i k}^{\tilde q}P_L){\tilde \chi}_k^{0}{\tilde q}_i+h.c.
\end{eqnarray}
 We also borrow the formulas for the partial decay widths of
$\tilde q_i$ ( $\tilde q_i$=$\tilde t_i$ and $\tilde b_i$) into  quark-chargino (or neutralino) from Ref. \cite{Bartl}:

\begin{eqnarray}\label{mkl22}
\Gamma(\tilde q_i\rightarrow q^{'}+\tilde
\chi_k^\pm)&=&\frac{g^2\lambda^{1/2}( m_{\tilde
q_{i}}^2,m_{q'}^{2},m_{\tilde \chi_k^\pm}^2)}{16\pi m_{\tilde
q_{i}}^3}\times \nonumber\\
&& {\left[{\left(|k_{i k}^{\tilde q}|^2+|\ell_{i k}^{\tilde
q}|^2\right)}(m_{\tilde q_{i}}^2-m_{q'}^{2}-m_{\tilde
\chi_k^\pm}^2)-4Re(k_{i k}^{\tilde q *}\ell_{i k}^{\tilde
q})m_{q'}m_{\tilde \chi_k^\pm}\right]}
\end{eqnarray}
and
\begin{eqnarray}\label{mkl23}
\Gamma(\tilde q_i\rightarrow q+\tilde
\chi_k^0)&=&\frac{g^2\lambda^{1/2}( m_{\tilde
q_{i}}^2,m_{q}^{2},m_{\tilde \chi_k^0}^2)}{16\pi m_{\tilde
q_{i}}^3}\times \nonumber\\
&& {\left[{\left(|a_{i k}^{\tilde q}|^2+|b_{i k}^{\tilde
q}|^2\right)}( m_{\tilde q_{i}}^2-m_{q}^{2}-m_{\tilde
\chi_k^0}^2)-4Re(a_{i k}^{\tilde q *}b_{i k}^{\tilde
q})m_{q}m_{\tilde \chi_k^0}\right]}
\end{eqnarray}
with $\lambda(x,y,z)=x^{2}+y^{2}+z^{2}-2(xy+xz+yz)$.

The explicit forms of $\ell_{i k}^{\tilde q}$, $k_{i k}^{\tilde q}$
and $a_{i k}^{\tilde q}$, $b_{i k}^{\tilde q}$ can be found in Ref.
\cite{Bartl}. We have to point out that although at the loop level
the SUSY-QCD corrections could be important, our analysis here are
merely at tree level, as can be seen from Eqs. (2.13) and (2.14). In
this work we content with tree-level amplitudes as we aim at
determining the phase-sensitivities of the decay rates, mainly.

\section{Top squark decays  }
Here we present the dependences of the $\tilde t_1$ and $\tilde t_2$
partial decay widths on $ \varphi_{U(1)}$ for $ \mu =200 $ GeV  and
for $ \mu =350 $ GeV. We also choose reasonable values for the
masses ($m_{\tilde t_1}$, $m_{\tilde t_2}$, $m_{\tilde \chi_1^\pm}$,
$m_{\tilde \chi_2^\pm}$, $m_{\tilde \chi_1^0}$) = (350 GeV, 800 GeV,
 180 GeV, 336 GeV, 150 GeV) for  $ \mu =200 $ GeV and ($m_{\tilde t_1}$, $m_{\tilde t_2}$, $m_{\tilde \chi_1^\pm}$,
$m_{\tilde \chi_2^\pm}$, $m_{\tilde \chi_1^0}$) = (350 GeV, 800 GeV,
 340 GeV, 680 GeV, 290 GeV) for $ \mu =350 $ GeV.

 For both sets of values by calculating the  $M_{\tilde
Q}$ and $M_{\tilde U}$ values corresponding to $m_{\tilde t_1}$ and
$m_{\tilde t_2}$, we plot the decay widths  for $M_{\tilde Q} \geq
M_{\tilde U}$  and $M_{\tilde Q} < M_{\tilde U}$, separately.
Fig.1(a) and Fig.1(b) show the partial decay widths of the channels
$ \tilde t_1\rightarrow b\tilde \chi_1^+
 $,  $ \tilde t_1\rightarrow b\tilde \chi_2^+ $   $ \tilde t_1\rightarrow t\tilde \chi_1^0 $,  $ \tilde t_2\rightarrow b\tilde \chi_1^+
 $,   $ \tilde t_2\rightarrow b\tilde \chi_2^+ $  and  $ \tilde t_2\rightarrow t\tilde \chi_1^0
 $   as a function of $ \varphi_{U(1)}$ for $ \mu =200$ GeV assuming $M_{\tilde Q} > M_{\tilde U}$ and
 $M_{\tilde Q} < M_{\tilde U}$, respectively. There we see some
 dependences on $ \varphi_{U(1)}$  phase. In order to see these dependences more
 pronouncedly, in Figs. 5(a)-(d) ($\tilde t_{1,2}$ decays for  $ \mu =200 $ GeV ) and Figs. 6(a)-(d) ($\tilde t_{1,2}$
 decays for  $ \mu =350 $ GeV ) we plot now separately only those decay channels
 whose dependences on $ \varphi_{U(1)}$ are not clearly seen in
 Fig.1 and Fig.2.
$ \Gamma( \tilde t_1\rightarrow t\tilde \chi_1^0 )$ and  $\Gamma(\tilde t_1\rightarrow b\tilde
 \chi_1^+)$ decay widths increase as $ \varphi_{U(1)}$ increases
 from 0 to $\pi$, but $\Gamma(\tilde t_1\rightarrow b\tilde
 \chi_2^+)$ width decreases as  $\varphi_{U(1)}$ increases. On the
 other hand, $ \Gamma( \tilde t_2\rightarrow t\tilde \chi_1^0 )$
 decrease for both  $M_{\tilde Q} > M_{\tilde U}$ and  $M_{\tilde Q} < M_{\tilde
 U}$ cases as  $\varphi_{U(1)}$ increases; $\Gamma(\tilde t_2\rightarrow b\tilde
 \chi_1^+)$ decreases for $M_{\tilde Q} > M_{\tilde U}$ but
 increases for $M_{\tilde Q} < M_{\tilde U}$ and $\Gamma(\tilde t_2\rightarrow b\tilde
 \chi_2^+)$ increases for $M_{\tilde Q} > M_{\tilde U}$ but
 decreases for $M_{\tilde Q} < M_{\tilde U}$.

 The branching ratios for $\tilde t_2$ are roughly $B(\tilde t_2\rightarrow b\tilde
 \chi_1^+)$ : $B(\tilde t_2\rightarrow t\tilde
 \chi_1^0)$ : $B(\tilde t_2\rightarrow b\tilde
 \chi_2^+)$ $\approx$ 8 : 2 : 1. This simply reflects both the large phase space
 and large Yukawa coupling for the decay $\tilde t_2\rightarrow b\tilde
 \chi_1^+$.

 Figs. 2(a) and 2(b) show the same partial decay widths for  $ \mu =350 $
 GeV. [See also Figs.6(a)-(d)]. They also reveal the significant
 dependences on CP-violation phase.

 For  $ \mu =350 $ GeV the WMAP-allowed band \cite{Belanger} takes place in larger $M_1$
 values ($ \sim 305-325$ GeV) leading to larger
 chargino and neutralino masses. This naturally leads very small
 decay widths for $\tilde t_1\rightarrow b\tilde
 \chi_1^+$ and kinematically forbidden $\tilde t_1\rightarrow b\tilde
 \chi_1^0$ and  $\tilde t_1\rightarrow b\tilde
 \chi_2^+$. Because of the large $\tilde t_2$ mass, $\tilde t_2$
 decay processes are still kinematically allowed as seen in Figs.
 2(a)-(b). The decay width of the process $\tilde t_2\rightarrow
 b\tilde \chi_1^+$ is the largest one among the $\tilde t_2$
 channels and the branching ratios are $B(\tilde t_2\rightarrow b\tilde
\chi_1^+)$ : $B(\tilde t_2\rightarrow b\tilde
\chi_2^+)$ : $B(\tilde t_2\rightarrow t\tilde
\chi_1^0)$ $\approx$ 4 : 2 : 0.3. The decay $\tilde t_2\rightarrow t\tilde\chi_1^{0}$ shows strong phase dependence.
\section{Bottom squark decays  }
We give sbottom decay widths as a function of  $ \varphi_{U(1)}$ in
Figs. 3(a)-(b)(for $ \mu =200 $ GeV ) and in Figs. 4(a)-(b)(for $
\mu =350 $ GeV ). [See also Figs. 7(a)-(d) and Figs. 8(a)-(c)]. Here
we choose the masses ($m_{\tilde b_1}$, $m_{\tilde b_2}$, $m_{\tilde
\chi_1^\pm}$, $m_{\tilde \chi_2^\pm}$, $m_{\tilde \chi_1^0}$) = (550
GeV, 800 GeV,
 180 GeV, 336 GeV, 150 GeV) for  $ \mu =200 $ GeV and ($m_{\tilde b_1}$, $m_{\tilde b_2}$, $m_{\tilde \chi_1^\pm}$,
$m_{\tilde \chi_2^\pm}$, $m_{\tilde \chi_1^0}$) = (550 GeV, 800 GeV,
 340 GeV, 680 GeV, 290 GeV) for $ \mu =350 $ GeV. $\Gamma(\tilde b_2\rightarrow b\tilde
 \chi_1^0)$ is smaller than $\Gamma(\tilde b_2\rightarrow t\tilde
 \chi_i^-)$ in spite of large phase space, because in $ \tilde b_2\rightarrow b\tilde \chi_1^0
 $ decay only $Y_b$ coupling enters. The dependences of the phase $
 \varphi_{U(1)}$ in sbottom decays are similar to those in stop
 decays.

 The branching ratios for $\tilde b_2$ decays are  $B(\tilde b_2\rightarrow t\tilde \chi_1^-)$ :
 $B(\tilde b_2\rightarrow t\tilde\chi_2^-)$ : $B(\tilde b_2\rightarrow
 b\tilde\chi_1^0)$  $\approx$ 10 : 7 : 0.5. While the process $\tilde b_2\rightarrow
 b\tilde\chi_1^0$ is suppressed more than one order its dependence
 on $\varphi_{U(1)}$ is prominent such that the value of decay
 width at  $\varphi_{U(1)}=0$ is 2 times larger than that at
 $\varphi_{U(1)}=\pi$.  $\varphi$-dependences of the process $\tilde b_2\rightarrow t\tilde
 \chi_1^-$ and $\tilde b_2\rightarrow t\tilde \chi_2^-$ can be
 seen easily in Fig. 3(a).

For $M_{\tilde Q}$$<$$M_{\tilde D}$, $\tilde b_1\rightarrow t\tilde \chi_1^-$ decay is about
five times larger than $\tilde b_2\rightarrow t\tilde \chi_1^-$. The branching ratios for $\tilde b_1$
decays are  $B(\tilde b_1\rightarrow t\tilde \chi_1^-)$ : $B(\tilde b_1\rightarrow t\tilde\chi_2^-)$ :
$B(\tilde b_1\rightarrow b\tilde\chi_1^0)$ $\approx$ 5 : 1 : 0.2.

\section{Conclusions  }
In this note, we have extended the study of third family squarks in
MSSM with complex parameters in Ref.\cite{Bartl} taking into account
the cosmological bounds imposed by WMAP data. For this purpose, we
have calculated numerically the decay widths of the third family
superpartners $\tilde t_{1,2}$ and $\tilde b_{1,2}$, in particular,
their dependences on the CP phase $ \varphi_{U(1)}$. We have found
that some decay channels like $\tilde t_2\rightarrow b\tilde
 \chi_1^+$,  $\tilde t_2\rightarrow b\tilde
 \chi_2^+$,  $\tilde t_2\rightarrow t\tilde
 \chi_1^0$, $\tilde t_1\rightarrow b\tilde
 \chi_1^+$, $\tilde b_1\rightarrow t\tilde
 \chi_1^-$, $\tilde b_1\rightarrow b\tilde
 \chi_1^0$, $\tilde b_2\rightarrow t\tilde
 \chi_1^-$ and $\tilde b_2\rightarrow t\tilde
 \chi_2^-$ show considerable dependences on
$ \varphi_{U(1)}$ phase. These decay modes will be observable at the
LHC. Therefore they provide viable probes of CP violation beyond the
simple CKM framework; moreover, they carry important information
about the mechanism that brakes Supersymmetry.

The HEP community eagerly waiting the discovery of superpartners at
the LHC. The obvious task in this context is to recognize them, for
example via their spins, experimentally.  There has been several
studies in this direction \cite{denegri,hisano}. Recently, Wang and
Yavin \cite{wang} have studied in detail the possibility of
measuring the spin of new particles in a variety of cascade decays,
especially in the decay channel  $\tilde q\rightarrow q\tilde
 \chi^\pm(\chi^0)$ which we consider here, and they conclude that the
prospects for spin determination are rather good.
\section*{Acknowledgment}
We would like to thank Durmus Ali Demir for suggesting the problem
and valuable comments.

\newpage
\begin{figure}
\includegraphics{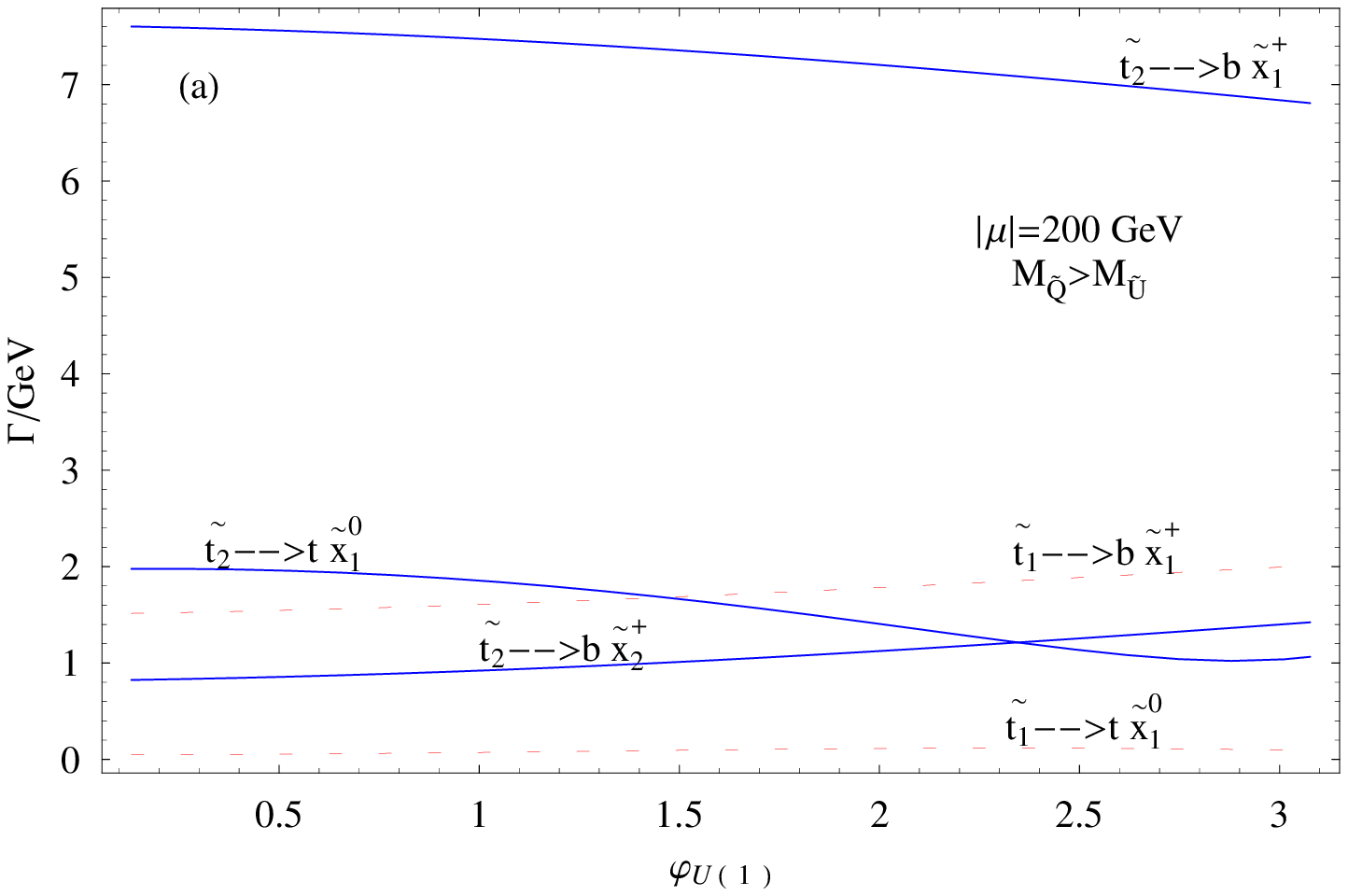} 
\label{figur1}
\end{figure}
\begin{figure}
\includegraphics{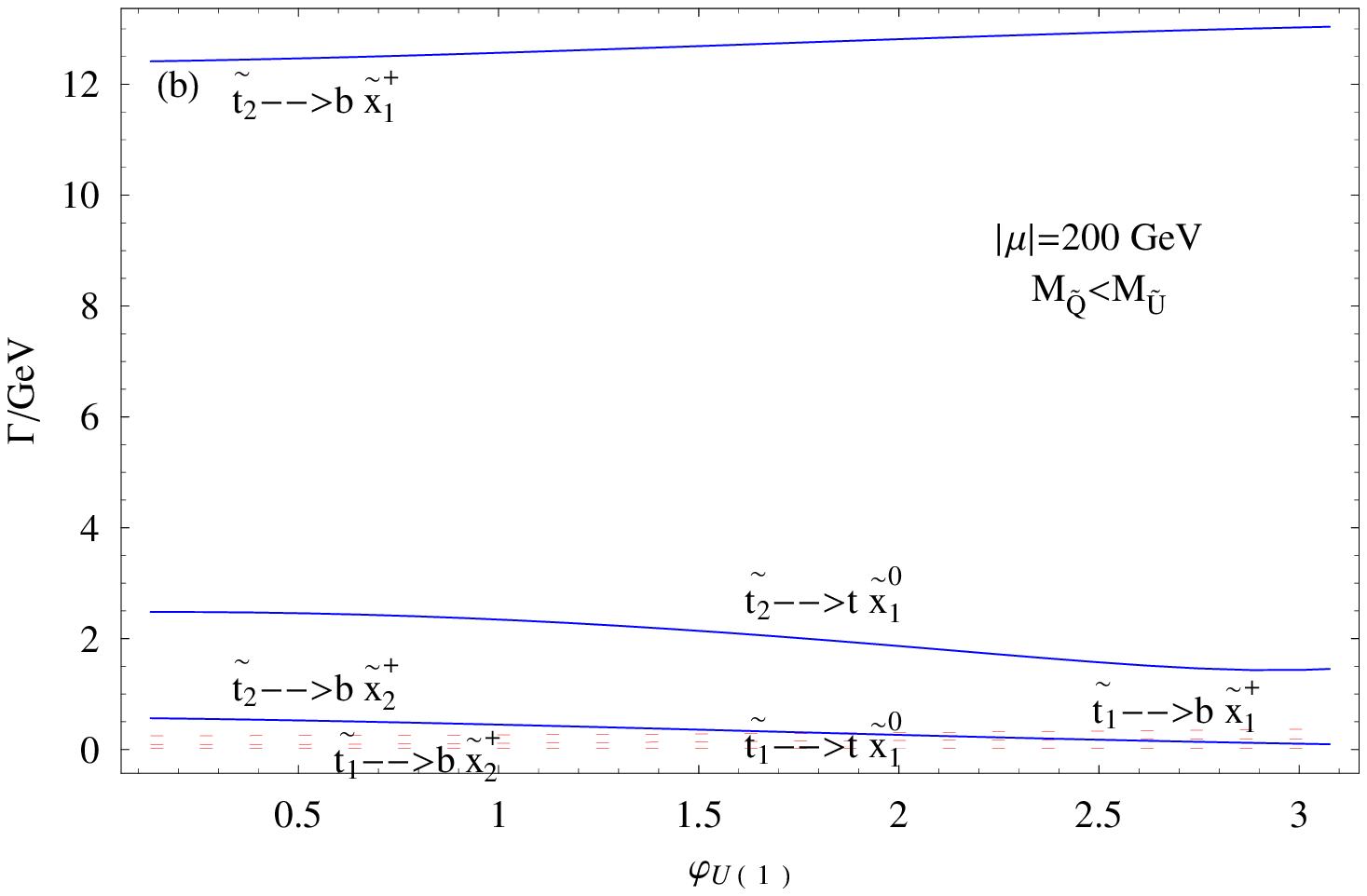} 
\caption{(a)-(b) Partial decay widths $\Gamma$ of the  $\tilde
t_{1,2}$ decays for $\mu =200 $ GeV , $\tan\beta=10$, $A_t=1.2 $
TeV, $\varphi_\mu$=$\varphi_{A_t}$=0,
  $m_{\tilde t_1}=350$ GeV and $m_{\tilde t_2}=800$ GeV. } \label{figur2}
\end{figure}
\begin{figure}
\includegraphics{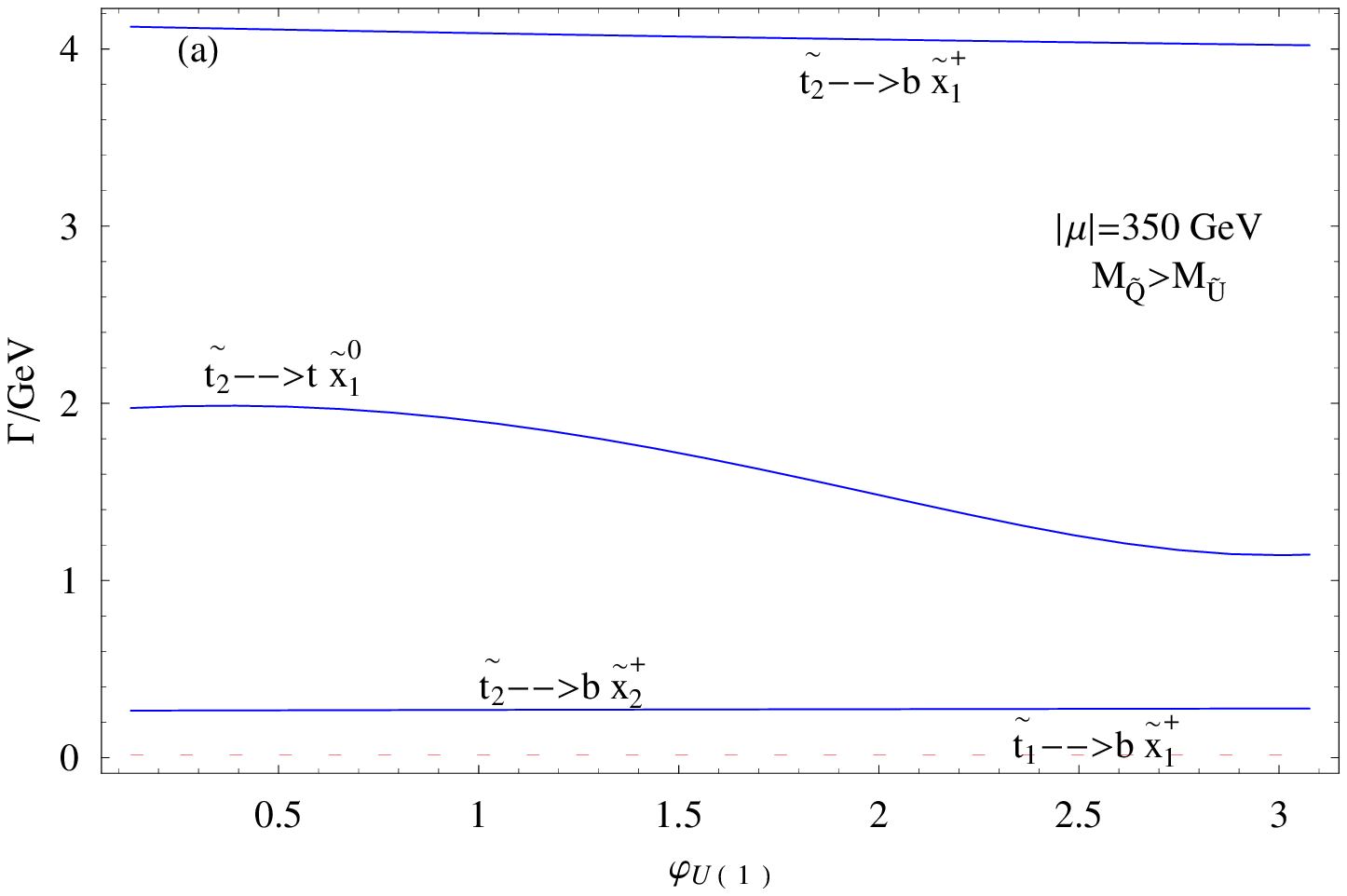} 
\label{figur3}
\end{figure}
\begin{figure}
\includegraphics{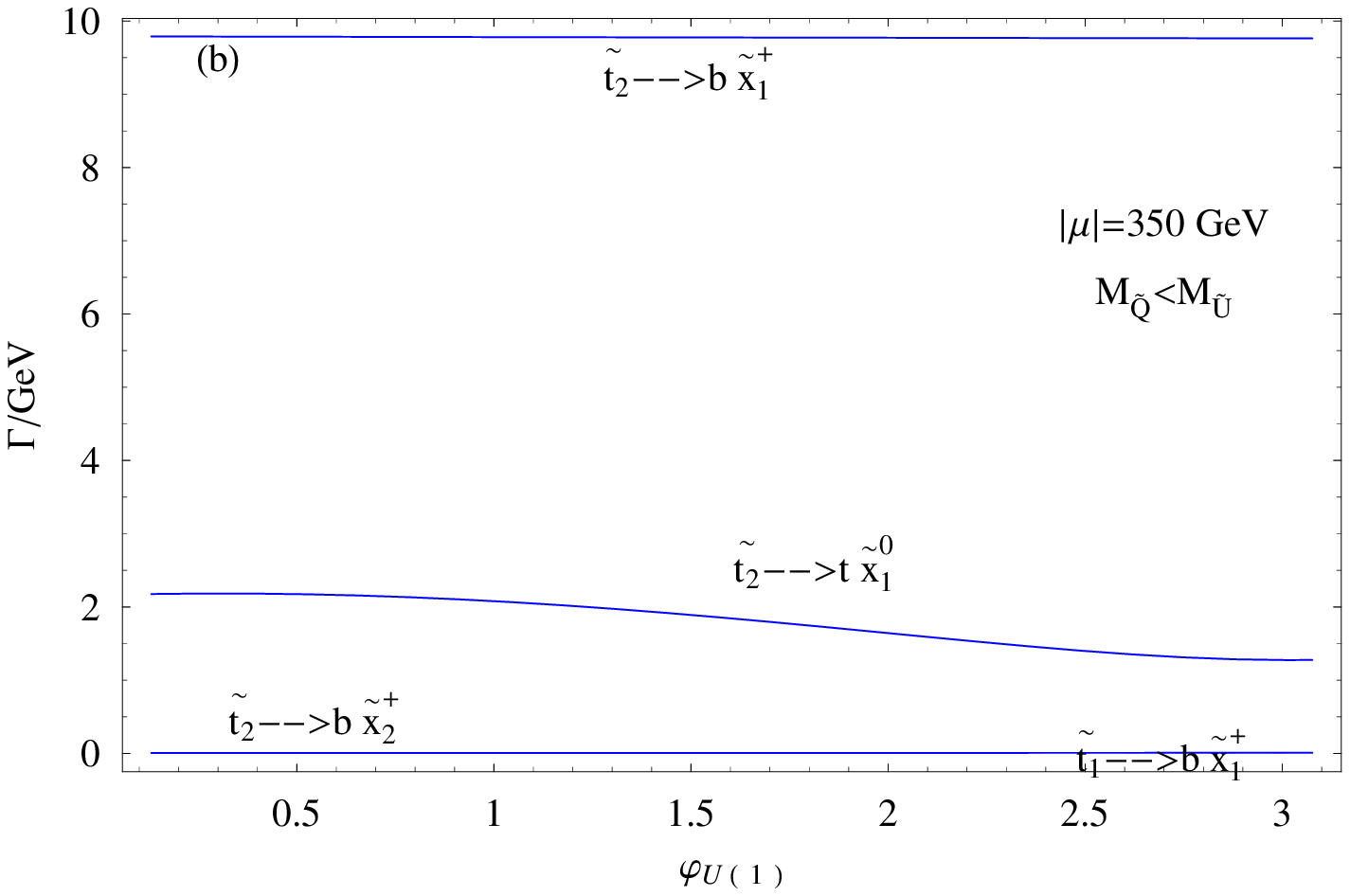} 
\caption{(a)-(b) Partial decay widths $\Gamma$ of the  $\tilde
t_{1,2}$ decays for $\mu =350 $ GeV , $\tan\beta=10$, $A_t=1.2 $
TeV, $\varphi_\mu$=$\varphi_{A_t}$=0
  $m_{\tilde t_1}=350$ GeV and $m_{\tilde t_2}=800$ GeV. }
\label{figur4}
\end{figure}
\begin{figure}
\includegraphics{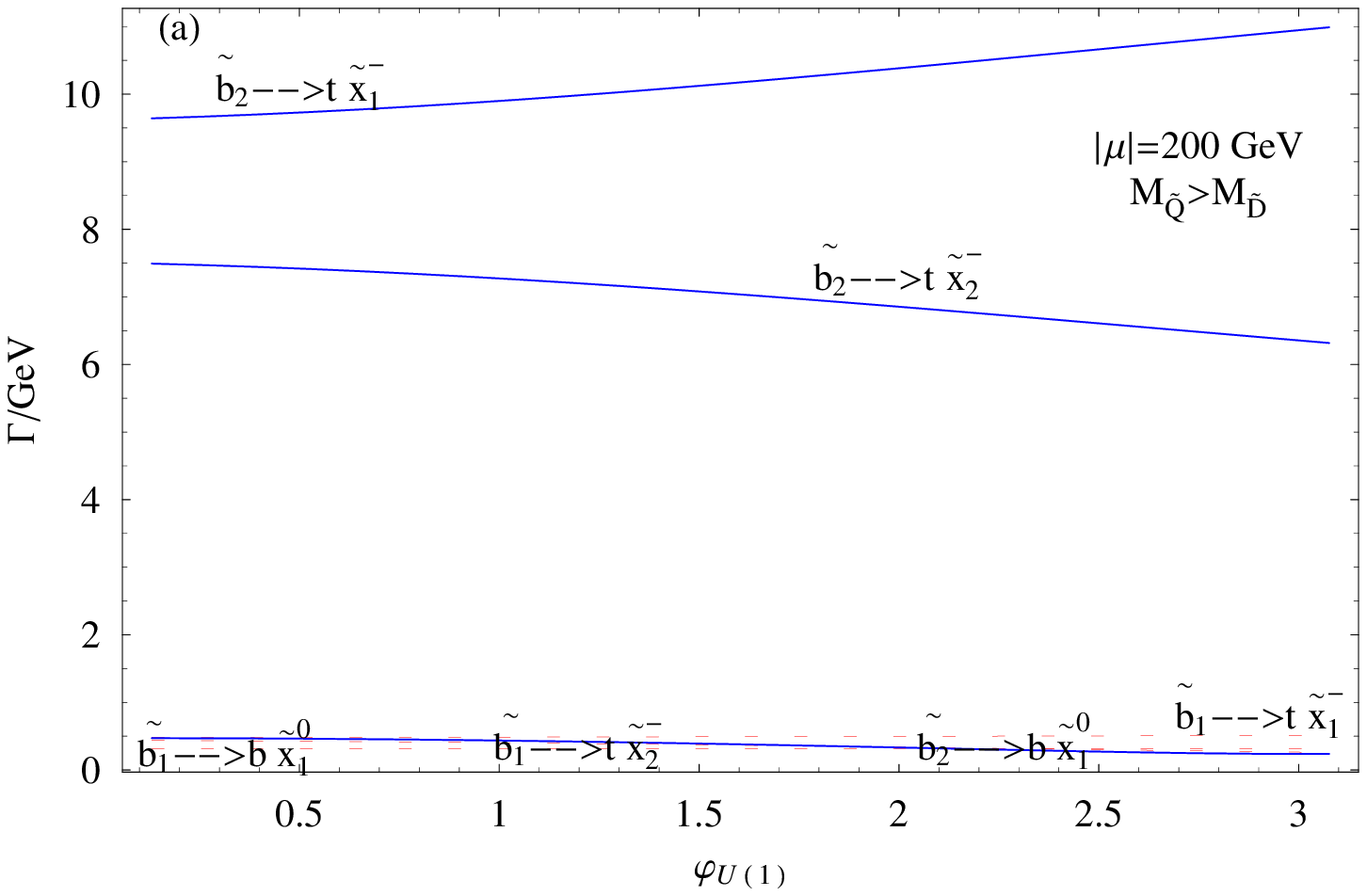} 
\label{figur5}
\end{figure}
\begin{figure}
\includegraphics{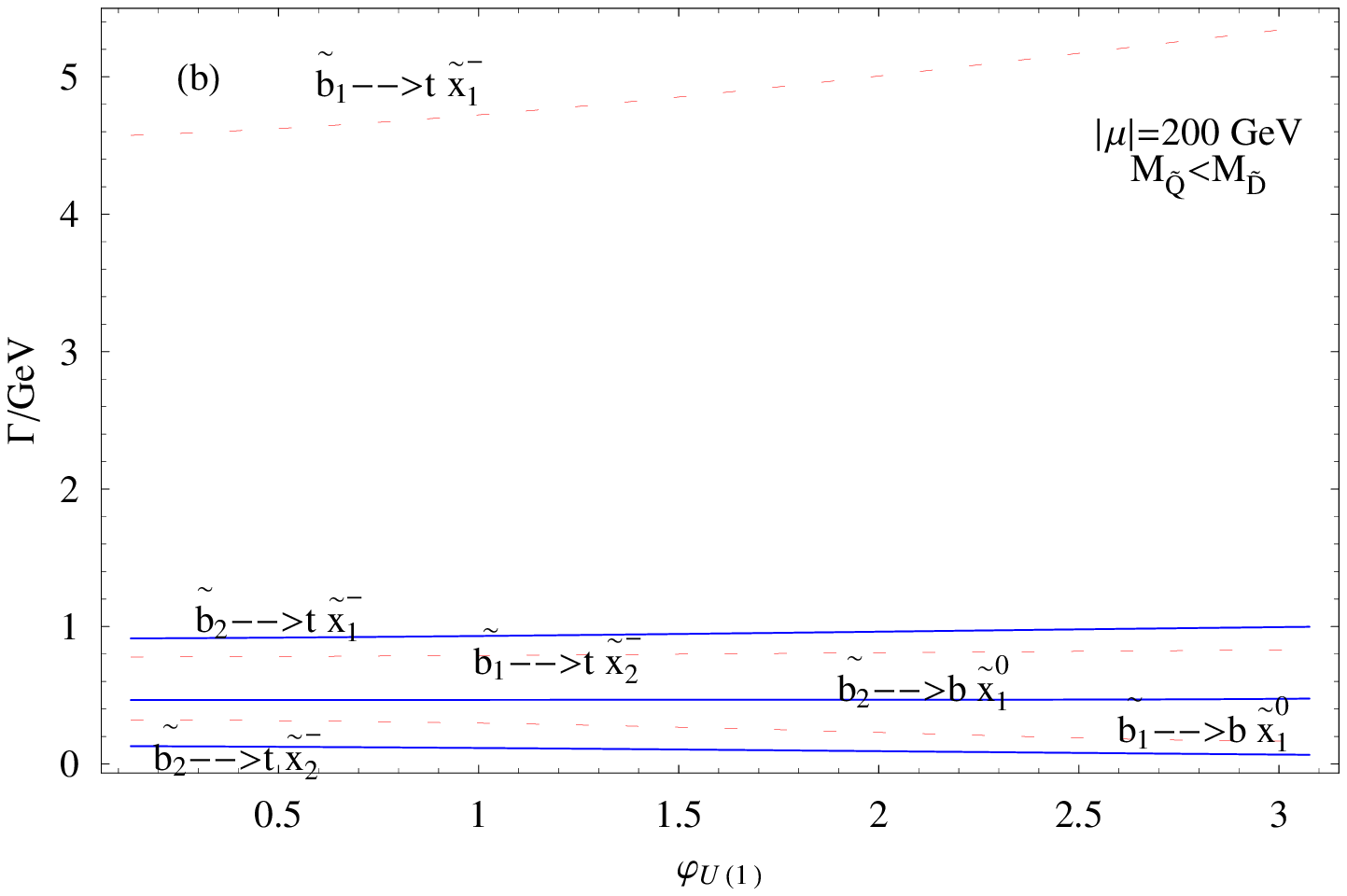} 
\caption{(a)-(b) Partial decay widths $\Gamma$ of the  $\tilde
b_{1,2}$ decays for $\mu =200 $ GeV , $\tan\beta=10$, $A_b=1.2 $
TeV, $\varphi_\mu$=$\varphi_{A_b}$=0 $m_{\tilde b_1}=550$ GeV and
$m_{\tilde b_2}=800$ GeV. } \label{figur6}
\end{figure}
\begin{figure}
\includegraphics{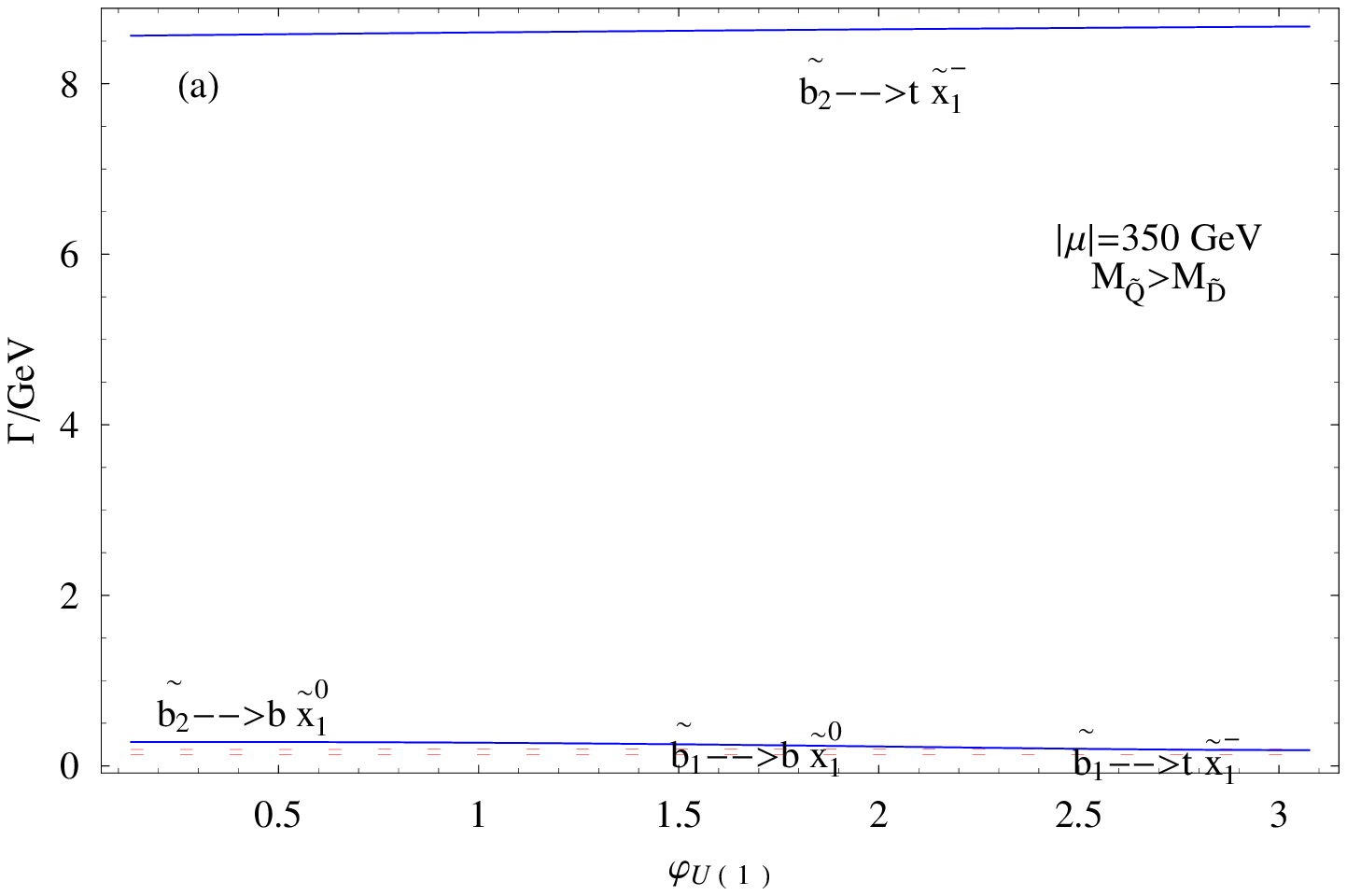} 
\label{figur7}
\end{figure}
\begin{figure}
\includegraphics{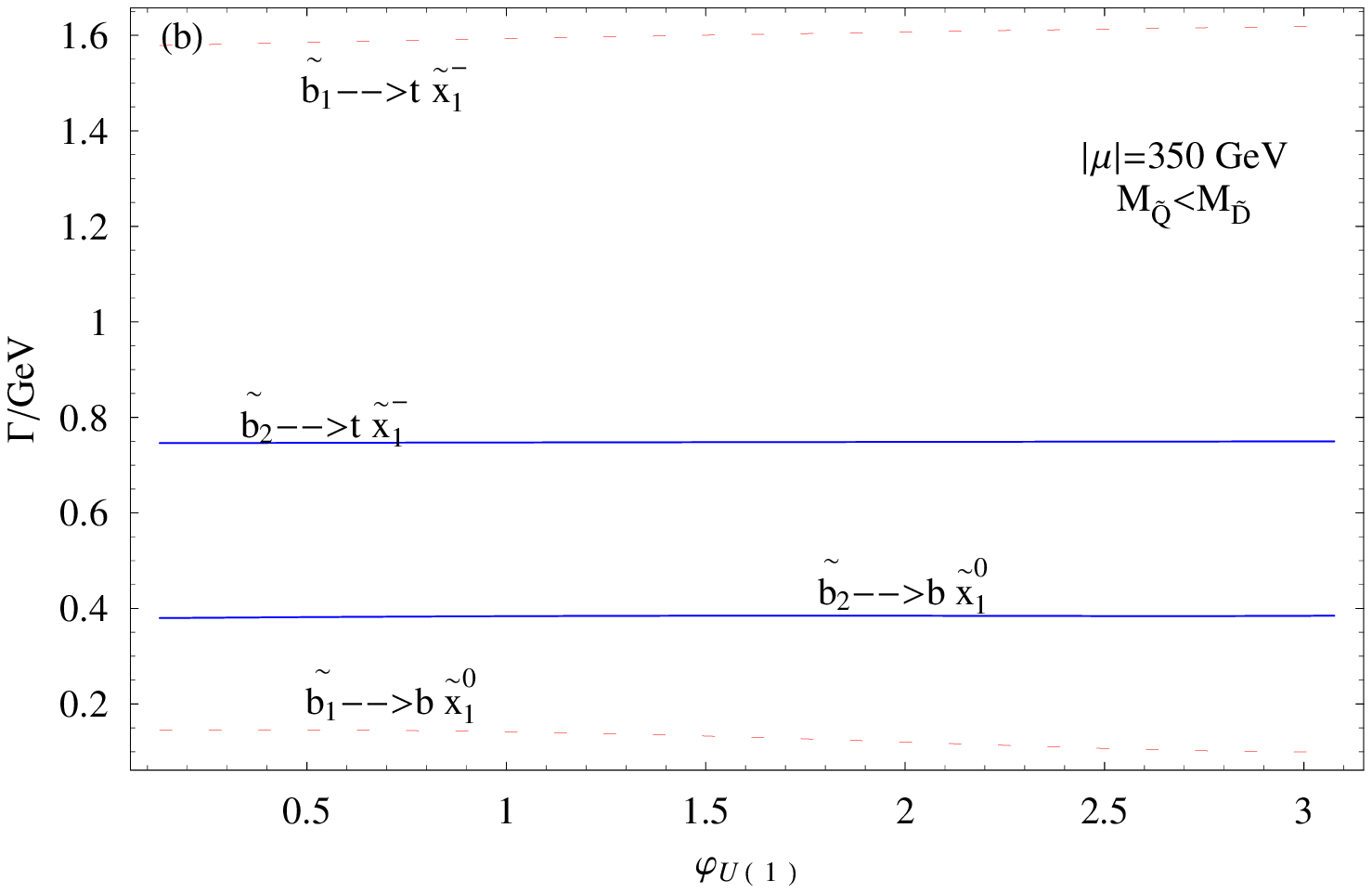} 
\caption{(a)-(b) Partial decay widths $\Gamma$ of the  $\tilde
b_{1,2}$ decays for $\mu =350 $ GeV , $\tan\beta=10$, $A_b=1.2 $
TeV, $\varphi_\mu$=$\varphi_{A_b}$=0 $m_{\tilde b_1}=550 $ GeV and
$m_{\tilde b_2}=800$ GeV. }
 \label{figur8}
\end{figure}
\begin{figure}
  \centering
\includegraphics[width=0.4\textwidth]{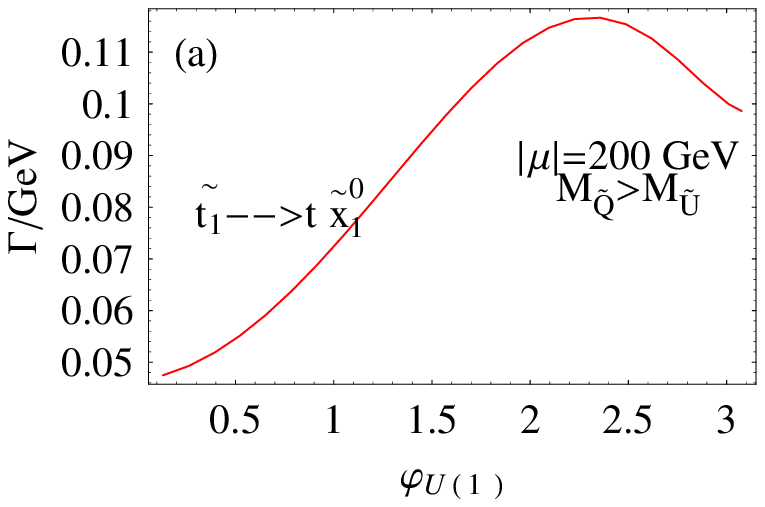}
\hspace{1cm}%
  \includegraphics[width=0.4\textwidth]{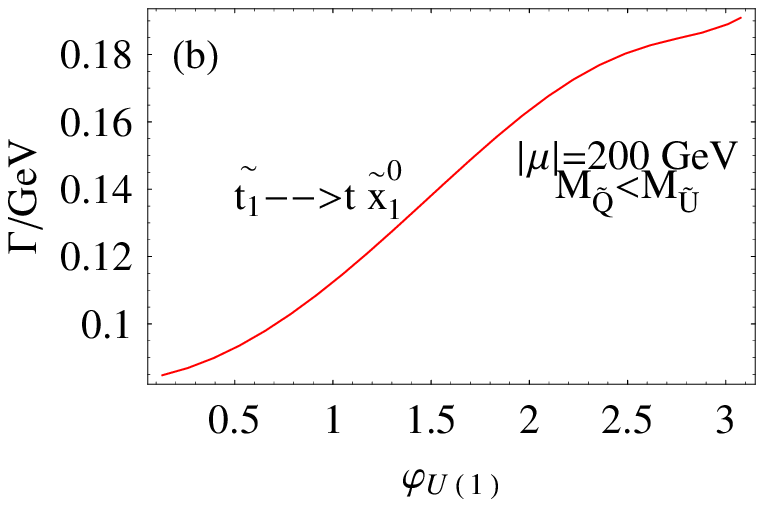}
  \label{dosfiguras1}
\end{figure}
\begin{figure}
  \centering
\includegraphics[width=0.4\textwidth]{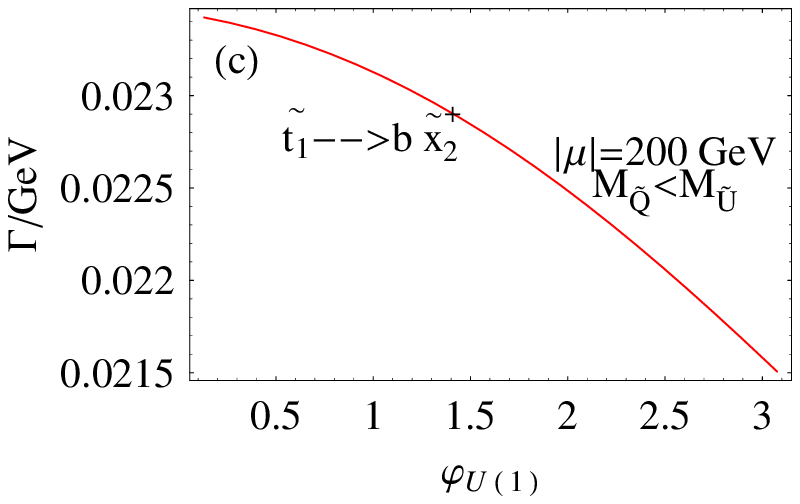}
\hspace{1cm}%
   \includegraphics[width=0.4\textwidth]{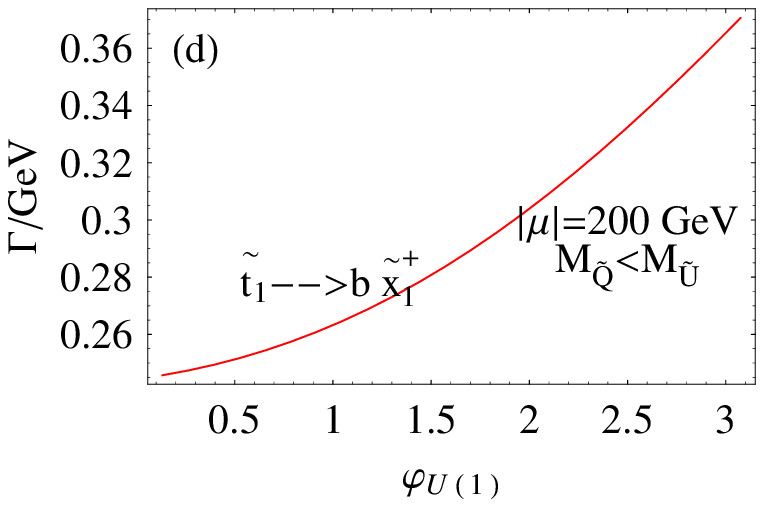}
\hspace{1cm}%
\caption{(a)-(d)  $ \varphi_{U(1)}$ dependences of certain $\tilde
t_{1,2}$ decays for $\mu =200$ GeV .}
  \label{dosfiguras2}
\end{figure}
\begin{figure}
  \centering
\includegraphics[width=0.4\textwidth]{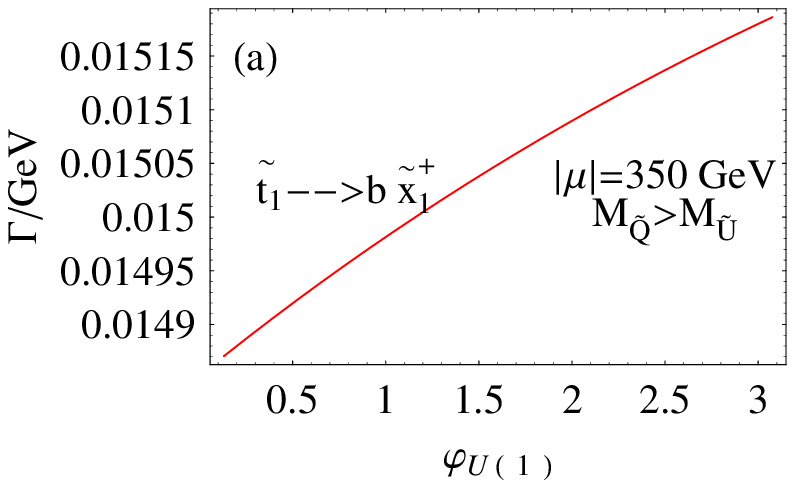}
\hspace{1cm}%
  \includegraphics[width=0.4\textwidth]{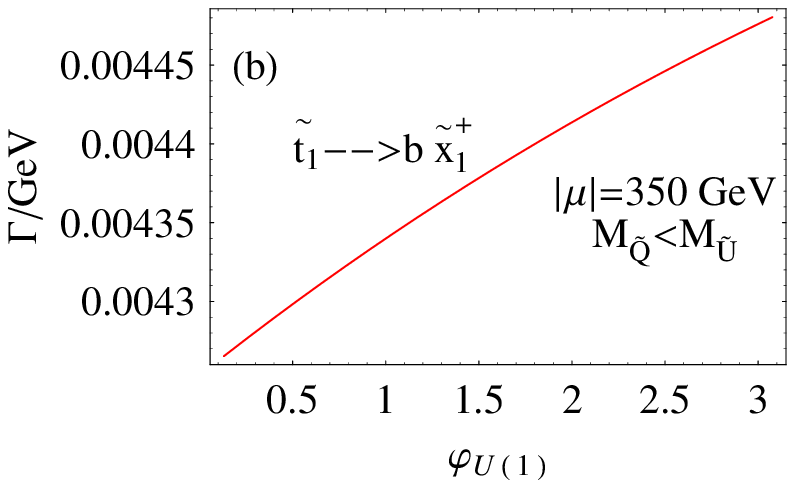}
  \label{dosfiguras4}
\end{figure}
\begin{figure}
  \centering
\includegraphics[width=0.4\textwidth]{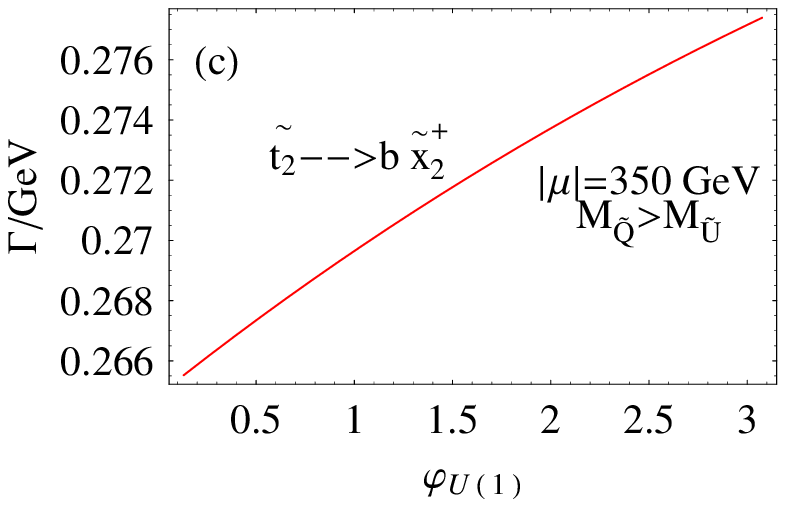}
 \hspace{1cm}%
  \includegraphics[width=0.4\textwidth]{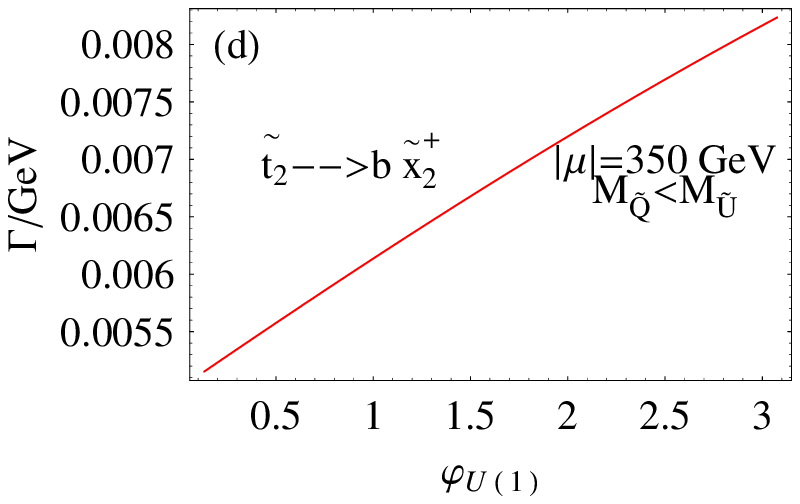}
  \caption{(a)-(d) $ \varphi_{U(1)}$ dependences of certain $\tilde t_{1,2}$ decays for $\mu =350$ GeV. }
  \label{dosfiguras6}
\end{figure}
\begin{figure}
  \centering
\includegraphics[width=0.4\textwidth]{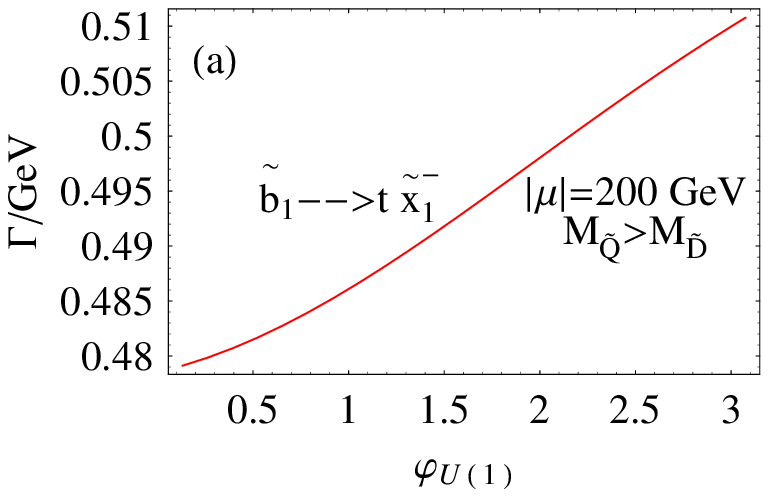}
\hspace{1cm}%
  \includegraphics[width=0.4\textwidth]{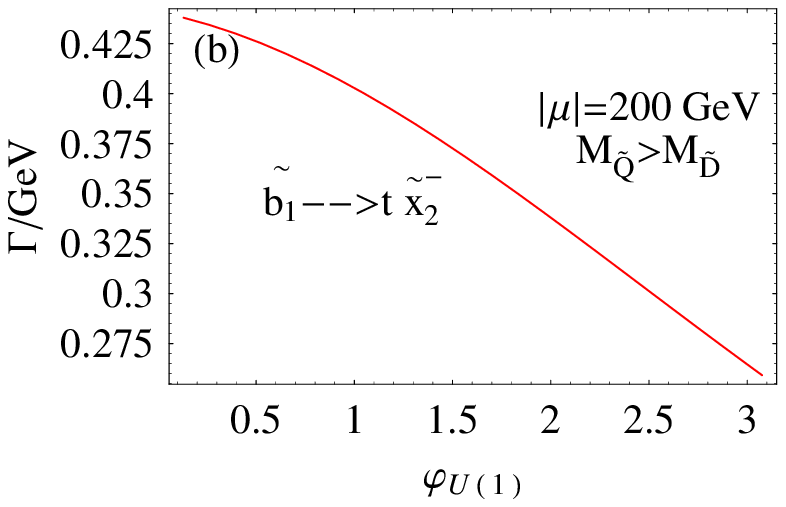}
  \label{dosfiguras4}
\end{figure}
\begin{figure}
  \centering
\includegraphics[width=0.4\textwidth]{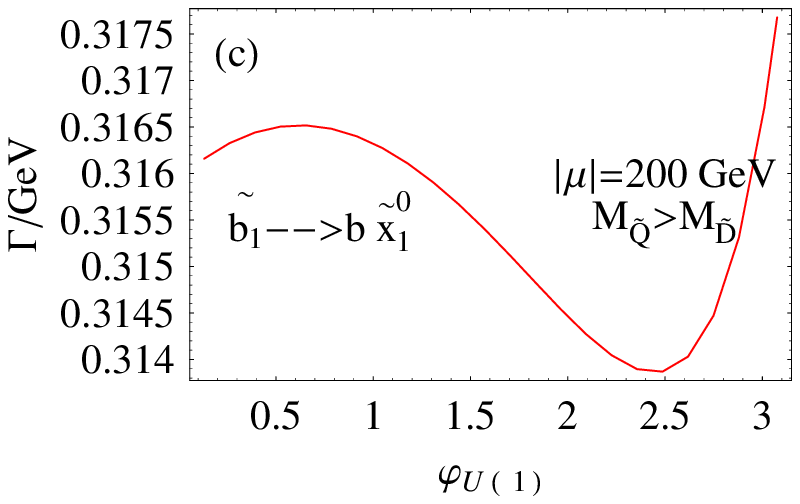}
 \hspace{1cm}%
  \includegraphics[width=0.4\textwidth]{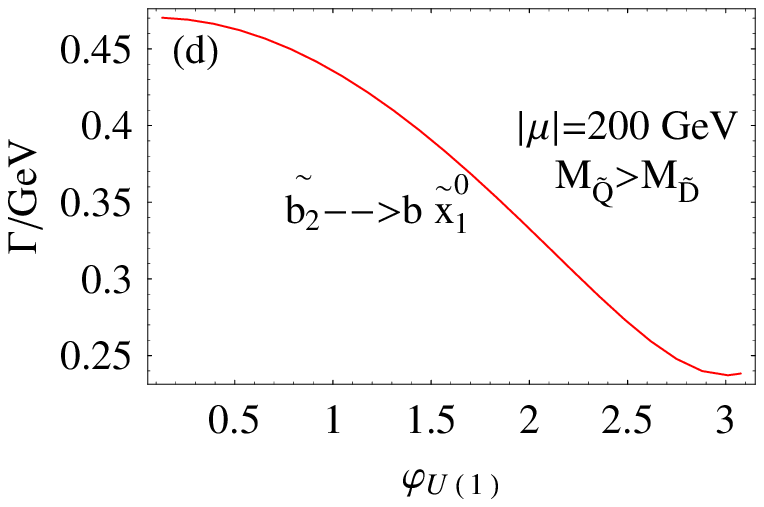}
  \caption{(a)-(d) $ \varphi_{U(1)}$ dependences of certain $\tilde b_{1,2}$ decays for $\mu =200$ GeV.}
  \label{dosfiguras6}
\end{figure}
\begin{figure}
  \centering
\includegraphics[width=0.4\textwidth]{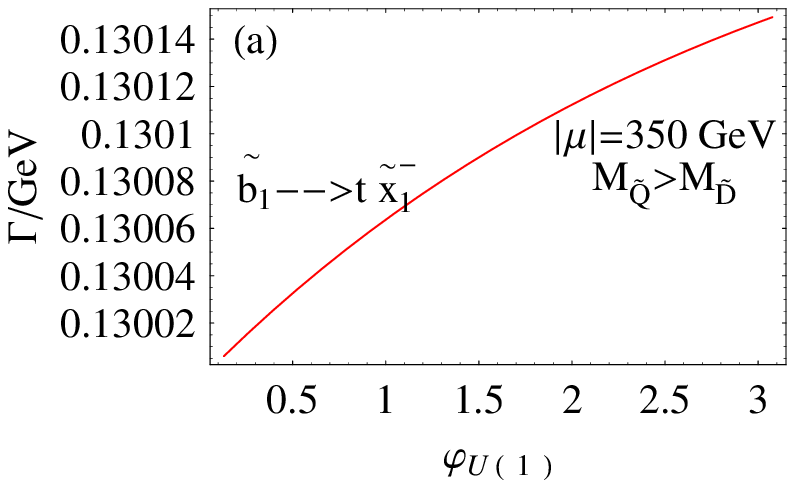}
\hspace{1cm}%
  \includegraphics[width=0.4\textwidth]{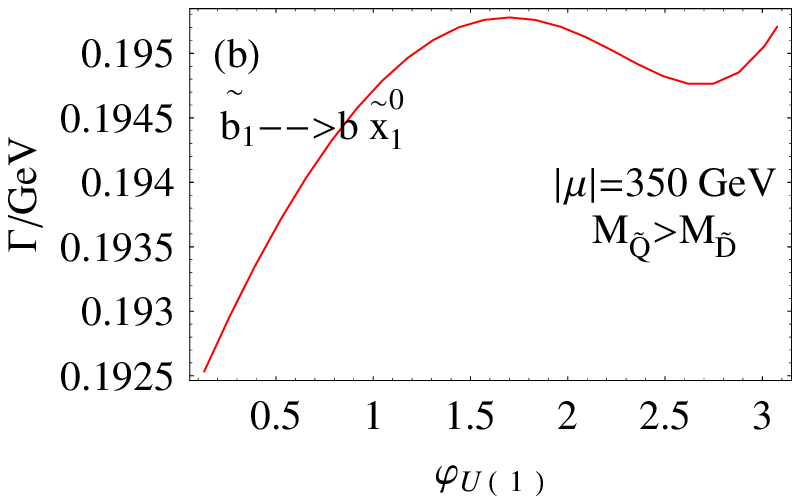}
  \label{dosfiguras8}
\end{figure}
\begin{figure}
  \centering
\includegraphics[width=0.4\textwidth]{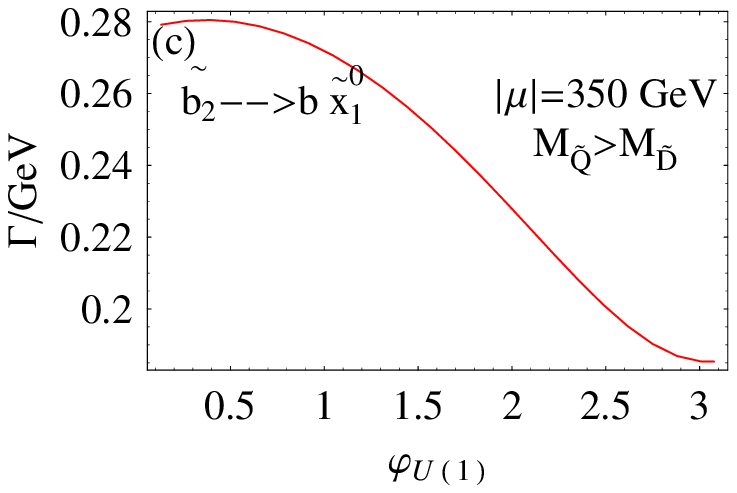}
 \caption{(a)-(d) $ \varphi_{U(1)}$ dependences of certain $\tilde b_{1,2}$ decays for $\mu =350$ GeV.}
  \label{dosfiguras10}
\end{figure}
\end{document}